\documentclass[a4paper,11pt,final]{article}
\usepackage{pos}

\usepackage{subcaption}
\usepackage{hyperref}
\usepackage{placeins}      %
\usepackage[sort&compress]{natbib}
\usepackage[notref,notcite]{showkeys}       %

\title{Over-abundant gamma-like signals around Solar disk shadows by  twin  bent  and smeared muon and  electron pairs  secondaries, versus rare  local TeV gamma}

\ShortTitle{Over-abundant gamma-like signals around Solar disk}

\author*[\orcid{0000-0003-3146-3932}, a,b]{Daniele Fargion}
\author[\orcid{0000-0002-9074-0584}, c,f]{Omar Tibolla}
\author[\orcid{0000-0001-7503-2064}, d]{Pier Giorgio De Sanctis Lucentini}
\author[\orcid{0000-0002-4603-8405}, e]{Sara Turriziani}
\author[c,f]{Sarah Kaufmann}
\author[g,h]{Danila Sopin}
\author[\orcid{0000-0002-1653-6964}, i]{Maxim Yu. Khlopov}

\affiliation[a]{Physics Department, Rome University 1,  Piazzale Aldo Moro 2, Rome, Italy}
\affiliation[b]{Osservatorio Astronomico di Capodimonte, INAF, Naples, Italy}
\affiliation[c]{Universidad Politecnica de Pachuca, Mexico}
\affiliation[d]{Physics Department, Gubkin University, Moscow, Russia}
\affiliation[e]{Centro de Astronomía (CITEVA), Universidad de Antofagasta, Av. Angamos 601, Antofagasta, Chile}
\affiliation[f]{University of Durham, United Kingdom}
\affiliation[g]{Institute of Physics, Southern Federal University, Stachki 194 Rostov on Don 344090, Russia}
\affiliation[h]{National Research Nuclear University MEPhI, 115409 Moscow, Russia}
\affiliation[i]{Virtual Institute of Astroparticle physics, 75018, Paris, France}

\emailAdd{daniele.fargion@fondazione.uniroma1.it}

\abstract{Cosmic rays with energies of tens of TeV and above, skimming the Sun, could fragment into pions. The resulting gamma photons and muons, as well as subsequent electron pairs, will reach us  in the form of gamma or electromagnetic air-showers , \textit{gamma-like}  air-showers  on Earth. Their multiple presence may soon be observed and disentangled by the LHAASO telescope array.}

\ConferenceLogo{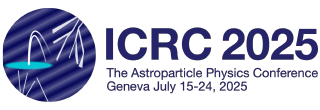}

\FullConference{39th International Cosmic Ray Conference (ICRC2025)\\
 15–24 July 2025\\
Geneva, Switzerland\\}

\graphicspath{{Figs/}{./}}
\def\orcid#1{\kern .08em\href{https://orcid.org/#1}{\includegraphics[keepaspectratio,width=0.6em]{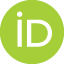}}}
\begin{document}
\maketitle

\section{Introduction: Skimming Cosmic Rays on solar atmosphere pointing to Earth}
\begin{figure}[b]
    \centering
    \begin{minipage}{0.50\textwidth}
        \centering
        \includegraphics[width=0.9\textwidth]{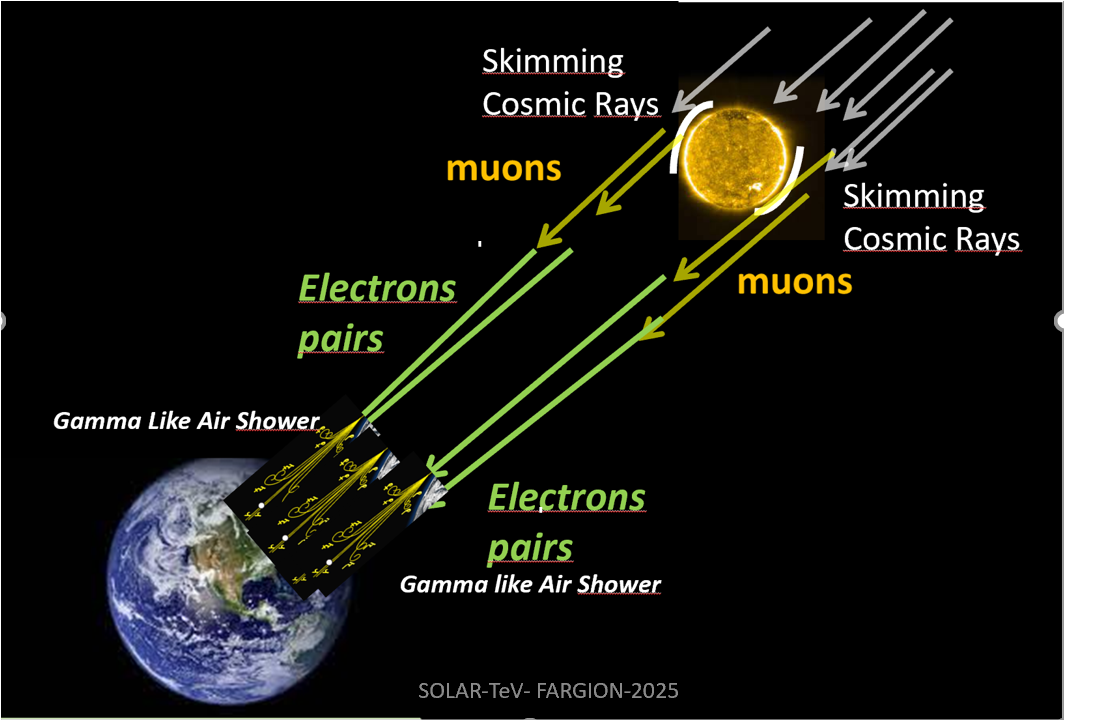} %
    \end{minipage}\hfill
    \begin{minipage}{0.45\textwidth}
        \centering
        \includegraphics[width=0.9\textwidth]{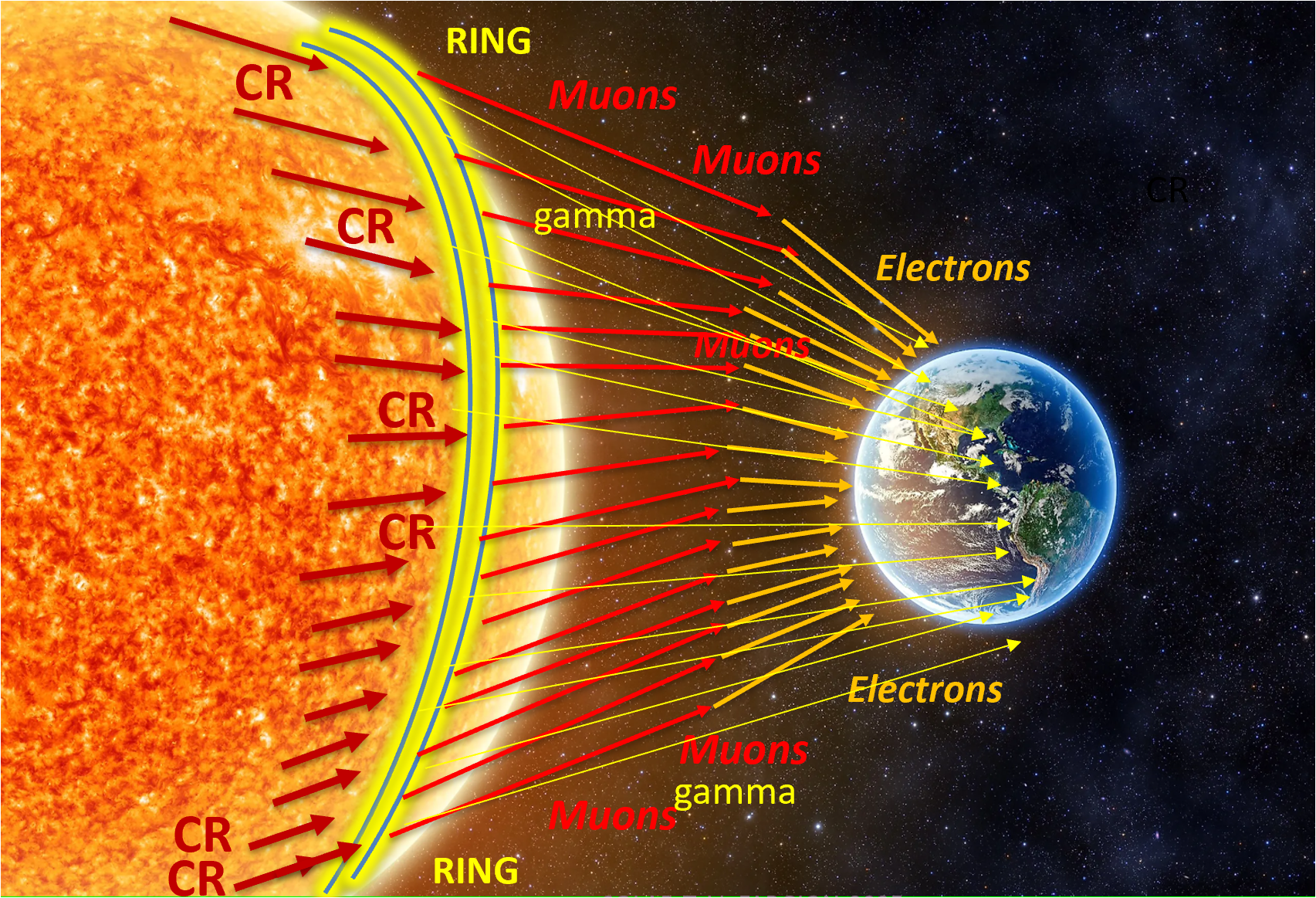} %
    \end{minipage}
        \caption{\textsc{Left.} A very schematic diagram showing PeV-TeV CR skimming the solar atmosphere heading toward Earth. Their secondaries are made by both neutral and charged pions. The neutral ones feed direct gamma-ray pairs along a thin ring-like layer of the solar atmosphere.  Such prompt gamma-ray  photons are often stopped by the solar atmosphere itself.  Charged pions and  their muons are much more penetrating, defining a deeper chord and a wider solar emission ring. Their beta decay secondaries at TeV may allow electron pairs to reach the Earth. These TeV electron pairs are behaving as any electromagnetic air-showers, producing a \textit{gamma-like}  air-shower  signals. 
        The final direct gamma-ray photons in thin layer could have been observed by HAWC \cite{albert2023discovery}. These events are also mimicked \textit{gamma-like}  air-shower  due to nearby TeV electron pairs hitting our sky near the same solar CR  shadow. In fact, HAWC observatory discovered a gamma ray excess within their Sun' shadow, which can be well described by  such  multiple shadow presences of  electromagnetic air-shower component.
        \textsc{Right.} Ideally, skimming CR and their secondaries, as above, define a thin layer or ring of the solar atmosphere from which the secondaries could emit TeV gamma or a wider ring of electron pairs, \textit{gamma-like}  air-shower   toward gamma ray  array on Earth.}
\label{fig:1}
\end{figure}

\begin{figure}[ht]
\centering
        \includegraphics[width=0.5\textwidth]{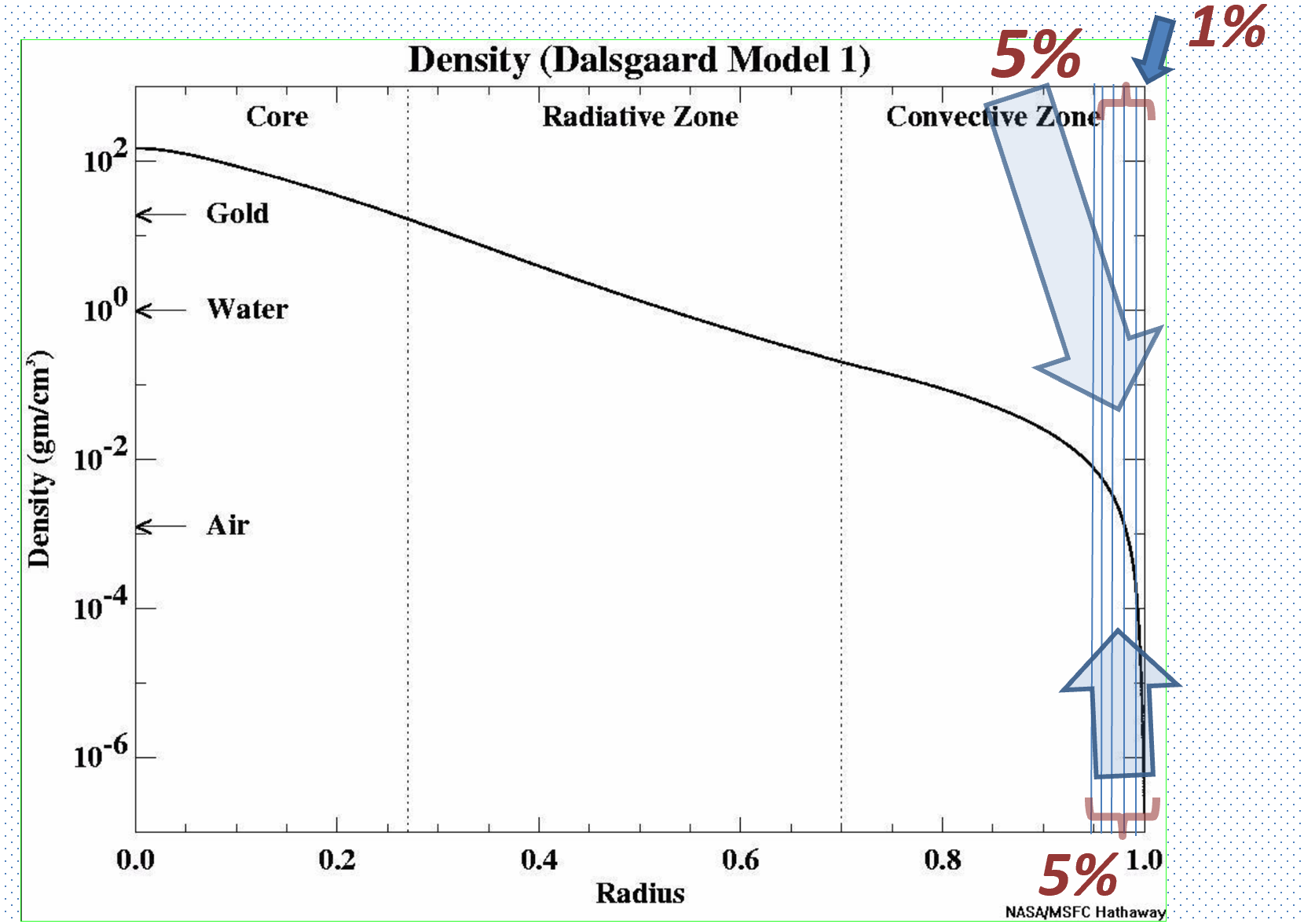} %
        \caption{Solar density vs radius. Depth required to explain the gamma-ray signal corresponds about to $5\%$ of the solar radius, that can be crossed , in principle , only  by  ideal penetrating 70 TeV muons. Such inner deeper chord  should cross in a  too dense  mass  ( or slant depth) to be overcome even for these energetic muon. Therefore tens TeV may cross only $1-2 \% $ of solar radius depth.
        The corresponding primary CR,  at hundreds of TeV or PeV , are too rare, in known and observed CR spectra, to explain the  over-abundant  HAWC data. Deflection (coherent or random)  of TeV muon pairs by a few degrees  within the local atmosphere solar field at TeV energy , inside granular solar plasma, could instead produce a compatible flux and  signals.  Image reprocessed in \cite{FargionICRC2023PoS1548} from Model S data published by Hathaway from NASA Marshall Space Flight Center.}
        \label{fig:4}
\end{figure}

Cosmic rays (CR) scattering on the Earth \cite{Grieder:2001ct}, on the planets and on the Sun atmosphere are a convenient laboratory for astrophysics. For instance, our Moon is the brightest source in the sky  about tens MeV gamma photons.   This occurs because dominant GeV CR hit a hard lunar soil target and their secondary pions, nearly at rest, produce gamma photons near tens MeV energy.  This is not taking place on soft atmosphere, because CR penetrate deeper  in the Sun  atmosphere and their vertical gammas are well absorbed.  But an Ultra High Energy (UHE) CR in the TeV-PeV  range,  skimming  on the solar  upper atmosphere horizons,  along shorter atmosphere chord, may have  secondaries that may  survive and exit, shining to us. It  is somewhat similar to  energetic PeV-EeV horizontal skimming tau neutrino or an up-going one from Earth \cite{fargion2004tau} leading first to an inner  neutrino  weak interaction  in the rock and,  soon  later , to a secondary tau, whose eventual  escape in air and  its decay lead to a  so called tau-air-shower \cite{fargion2002discovering}  (often called skimming neutrino). 
Therefore, CR  skimming our diluted horizontal atmosphere may lead to a long beamed air-shower secondary signals as penetrating hundred GeV muon at horizons . The  top solar atmosphere may also offer such a target for  skimming CR.  Indeed, random CR with energies from TeV  and above, while skimming the Sun, could interact and emerge aligned towards the Earth (see figure~\ref{fig:1}, left).
Their interactions within the solar atmosphere may produce TeV secondaries as neutral and charged pions or Kaons. The neutral pions feed soon twin gamma rays along a very thin solar surface, that will be observable on Earth as gamma-ray air-showers.  The charged  pions feed the more penetrating muons, whose later secondaries TeV-PeV electron pairs may also reach the Earth, leading to electron pair that produce electromagnetic showers. We shall call them \textit{gamma-like} air-showers (see figure \ref{fig:1}, right). Several TeV electron secondaries, while spiraling on vertical solar magnetic fields, may also hit solar photons feeding TeV gammas by Inverse Compton Scattering (ICS) \cite{fargion1998inverse}. 
Even electrons of tens of TeV can partially emit gamma rays by bremsstrahlung when they exit the solar atmosphere.
Their overall local gamma  ray signals should form a very thin ring of photons at Sun edge. An additional TeV signal made by positive-negative ring (or belt) of electron pairs is deflected by interplanetary solar field around the CR solar shadows. 
Indeed, skimming muon pairs at TeV can penetrate deeper and longer chords before escaping the Sun and heading towards Earth, undergoing beta decay in flight.
Their electron pairs might be also bent by interplanetary solar field and hit at the end the terrestrial atmosphere, behaving at TeV energies, as a common gamma ray electromagnetic air-showers (we shall call them \textit{gamma-like}  air-showers). 
The bending by the solar interplanetary magnetic spiral fields during their travel will split their contribute to a twin gamma-like spot or ring, a positive and a negative one, nearly one-two degree each apart, at TeV energy.  The gamma-ray excess discovered by HAWC within its Sun' shadow in 2023 might be compatible with this two additional shadow scenario~\cite{albert2023discovery}.
The recently expanded LHAASO gamma ray array area, given its more abundant statistics, should soon be able to distinguish negative from positive spots with respect to the central un-bent neutral component;
this latter has to be, in our present expectation, a minor complementary TeV signal.  The major contribute to the twin gamma-ray-like spots could be made by originated skimming muons, locally smeared inside the solar chromosphere by solar fields,  and their later electrons pairs, partially  bent in their flight by interplanetary magnetic fields.   At TeV energy and above  CR protons or electrons are  only 1 or 2 degrees deviated by the interplanetary system magnetic fields. Therefore, above 1$\sim$TeV, the solar CR proton shadow   is  quite well localized and only a little bent and smeared, offering an ideal calibration for the array angular resolution. This test had been applied by past and recent array detectors, as ARGO, HAWC and last LHAASO.
The secondaries of skimming  TeV CR  may lead to neutral pions and their prompt photons at half energy. These photons may hardly reach us from their  thin layer of production, being mostly soon absorbed. The same  secondaries skimming electron pairs  may also shine in gamma-ray-rays by bremsstrahlung  or by ICS on thermal solar photons. Such ICS could be  efficient  if electron pairs may stand a while in spirals along solar magnetic fields.  The charged CR secondaries, pions and kaons, may freely fly in  horizontal solar diluted atmosphere $(\rho = 10^ {-6} \cdot g\cdot cm^{-3})$, only for few tens  of kilometers  because:
$$ L_\pi = c\cdot \tau_{\pi} \cdot \gamma_{\pi} = 56 (E_{\pi} /TeV)\cdot   km $$ 
with almost negligible  (one every twenty)  hadron interactions. Indeed, the corresponding proton-proton  interaction length is longer:
 $$ L_{(pp)}  = 1/ (\sigma_i \cdot N_A \cdot \rho) = 560 km (10^{-6} g \cdot (cm^{-3} /\rho))$$.
Moreover, considering the pion interaction on the diluted solar plasma, its interaction survival length is almost: 
$ L_{(i,p-pion)} = 1/ (\sigma_i \cdot N_A \cdot \rho) = 1100\cdot (10^{-6}g \cdot cm^{-3}/\rho))\cdot km $,  allowing a quite successful, energetic TeV muon decay.  These penetrating  leptons may fly at much deeper distance  as long as  $$L_\mu = c\cdot \tau_\mu \cdot \gamma _\mu = 6,23 \cdot 10^3 \cdot (E_\mu /TeV) \cdot km $$
in the solar atmosphere, even in a more deep and dense ($\rho= 10^{-4} \cdot g \cdot cm^{-3}$)  inner chords.  These skimming TeV muons may often escape and flight later to the Earth: soon they may decay into  TeV electron pairs that may reach and hit finally the terrestrial atmosphere, producing  an electromagnetic cascade or a gamma-like air-shower at TeV energy.  They will be just arriving  (with bending due to interplanetary magnetic fields) along the same solar shadow disk, at their edges:  the TeV  positron will deflect and  falls as most of TeV CR  proton at same energy,  while the negative one will reach Earth  on the opposite solar  CR  shadow corner.  The  TeV electron pairs may reach Earth producing an electromagnetic air-showers, thus producing more abundant  \textit{\textit{gamma-like} }  signals in a smeared dipolar asymmetry.  The central undeflected gamma-ray spot might be an additional central minor component.
It is interesting to recall that the interaction of TeV gamma-ray photons on air in the Earth's atmosphere has a characteristic length shorter (almost half) than the interaction length of hadrons.  Indeed the TeV photon  cross-section  is twice as large as that of the proton-air interaction at same energy.
This is mostly due to the large gamma  production of electron pairs  onto air Nitrogen and Oxygen atoms, an effect due to the Coulomb fields (Bethe-Heitler scenario) amplified by $Z^2$ nuclei term. 
However, we underline  here that the solar atmosphere is dominated mostly  by proton nuclei and not by any light nuclei as on Earth atmosphere.  Therefore Z=1 offer a negligle amplification. This reduces, surprisingly,  the TeV gamma-ray interaction cross-section on the Sun atmosphere to a fairly small $\sigma_\gamma = 9 \cdot mb$, with respect to a larger inelastic hadron one at TeV energy ($\sigma_p  =   30 mb$).  This implies that  the gamma-ray TeV photons may penetrate (noot half as on air) but  as much as 3.3 times deeper than the hadron distances, or nearly $L_\gamma = 1850 \cdot km$ at the lowest solar density atmosphere $(10^{-6} g \cdot (cm^{-3} /\rho))$.  Assuming this gamma-ray ideal distance, their corresponding solar height altitude h, part of the ring size, by geometry,  is  only $h= 8.8\cdot 10^{-7} \cdot R_{\odot}$ or  a corresponding thin ring size $2 \cdot \pi \cdot h \cdot  R_{\odot} $ and its area of  $ 1.76\cdot 10^{-6} $ of the solar disk. A very thin area and a negligible gamma-ray emission power.
Further relativistic phenomena, as we will show, can increase the opening angle of the gamma-ray emission, transforming it into a larger ring. But not yet, as large as the  needed one to explain the HAWC over-abundant  data. See the solar  atmosphere profile and density, in figure  \ref{fig:4}: it defines the TeV muon-electron pairs and the eventual gamma-ray production place.

\section{Neutral pions decay into TeV photons, their opacity  forming a thin ring}
As mentioned above, tens TeV CR skimming the solar atmosphere will hit mostly proton targets and a quite smaller amount  of Helium nuclei, at rest in the Sun photosphere.
In  the Sun photosphere, the CR above $\sim$10 TeV skimming solar atmosphere hit mostly target protons and a few Helium nuclei. Therefore, the UHE proton-proton deep inelastic scattering will be in a quark-quark regime, forming a cascade of tens of secondaries, most of which will be charged pions and  kaons (one third will be neutral ones). The $\pi^0$ prompt decay into photon pairs  could be the first channel of  TeV photons production. However, their same electromagnetic  cross-section  will lead to the new born photons scattering soon  on the same solar proton atmosphere target. Such an hadron and electromagnetic bounding opacity  will occur  even in diluted solar atmosphere. Only a very tangent, up-going external photons will escape. This imply, as mentioned,  a very thin layer of the Sun for such  prompt TeV gamma-ray signal, see figure \ref{fig:1}. To make a first   scenario to reach such a TeV gamma-ray, we might assume an UHE  CR proton  around a  8-10 TeV  energy,  whose  leading particle  $\pi^0$  will be at 2 TeV energy, with  a wide ($18 \pm 4$) number of charged pions at hundreds GeV energy, a third of them in neutral $\pi^0$ form, a  few   Kaons, and a few  baryons,  mostly nucleons at several hundred GeV energy, with additional tail of resonances and neutrinos.

Let us  re-consider as in  figure \ref{fig:4} the first most diluted  high altitude  of solar convective zone at 
density $ \rho = 10 ^{-7} \cdot g \cdot cm^{-3} $.   The   hadron cross section  with protons defines   common a survival distance lenght L of nearly  $L = 560~km$,  for gamma-ray  or electron pair production  at longest distance  $D = 1.850 \cdot 10^3 \cdot  km $  at   $\rho = 10^{-6}  \cdot g \cdot cm^{-3} $. 
As we mentioned above such distance chord D, even at such diluted atmosphere, the height below the horizons $h$ is derived, within one percent accuracy: $h=  D^2 / (8\cdot R_{\odot})$, where $R_{\odot}$ is the solar radius.    At  a density  $\rho = 10^{-7}  \cdot g \cdot cm^{-3} $ the result is  only $\delta A_{\odot} = 2 \cdot 10^{-4} \cdot  A_{\odot}$ , The  emission by  such a  thin ring is still negligible. The secondary electron pairs are also skimming with the same narrow distances and  inside a thin area, but we shall soon show,  they may be bent locally and smeared along their flight.  However, there is  finally an interesting additional relativistic argument for a   wider spread of the  prompt gamma-rays pairs.

\section{ Relativistic split of neutral and charged pions  and their decays }
During the (8-10) TeV CR proton interaction with a solar proton target, the opening angle of the twin secondary fragments is, at first approximation, just  $\Theta = 0.57^o$ degree. This offer to pion and gamma pairs  a non negligible  ring size   $(2\cdot \theta/\pi \cdot R_{\odot}= 0.6\% R_{\odot}$)  and a  corresponding fraction of solar disk area being as large as $\delta A_{\ odot} = 1.2\% A_{\odot}$. Anyway this estimation cannot explain all the observed over-abundance gamma-like  signals, which is reaching the Earth from nearly  $10\%$ of the solar disk area.
The  most probable secondary signals on Earth by skimming CR  on solar atmosphere are those by charged pions and kaons mesons:   they may flight as noted above, by nearly  $(50 /((\rho/10^{-5} ) \cdot g \cdot (cm^{-3} ))$ km.   
However, their secondary muons are even  much more  penetrating  even while being  
at deeper  and more dense regions. Muons might run as  much as  distances $L_\mu = 6.6 \ 10^3 km (E_\mu /TeV)$ even around  $\rho = 1,5 \cdot 10^{-4}  \cdot g \cdot cm^{-3} $   within solar atmosphere density .  Indeed  
these muon pairs may be  crossing very long chord inside the solar atmosphere. Several tens TeV muons  may penetrate deep height and widest chord, as deep as  one or even almost two percent of the solar radius. The corresponding chord distance and muon energy are respectively:
$$h =1\% \cdot R_{\odot} ; D= 1.97\cdot 10^5 km;  E_\mu= 30~TeV$$ 
$$h=2\% R_{\odot} ; D= 2.8 \cdot 10^5 km;  E_\mu= 44~TeV.$$ 
Their emission area is correspondingly at $2\% $ or $4\% $ of the solar disk area. The different charge deflection may double the ring area to $4\% $  and $8\% $. Indeed, their in-flight beta decay may lead to electron pairs at tens or several TeV energies, whose electromagnetic  air-showers  on Earth would  mimic the gamma-ray ones in any ground array as the gamma-like events observed by HAWC.  At deeper height ($h \ge 3\%~\cdot R_{\odot}$)  and  corresponding  wider chords ($D= 3.4\cdot 10^5 km$), the denser atmosphere  to be encountered  reduces or absorb the muon  energy and  nearly avoid their crossing.   The present knowledge of CR spectra at tens TeV in space, assumed comparable on the Sun,  is too low to be consistent with the over-abundant  HAWC data. 

\begin{figure}[tb]
    \centering
    \begin{minipage}{0.45\textwidth}
        \centering
        \includegraphics[width=0.9\textwidth]{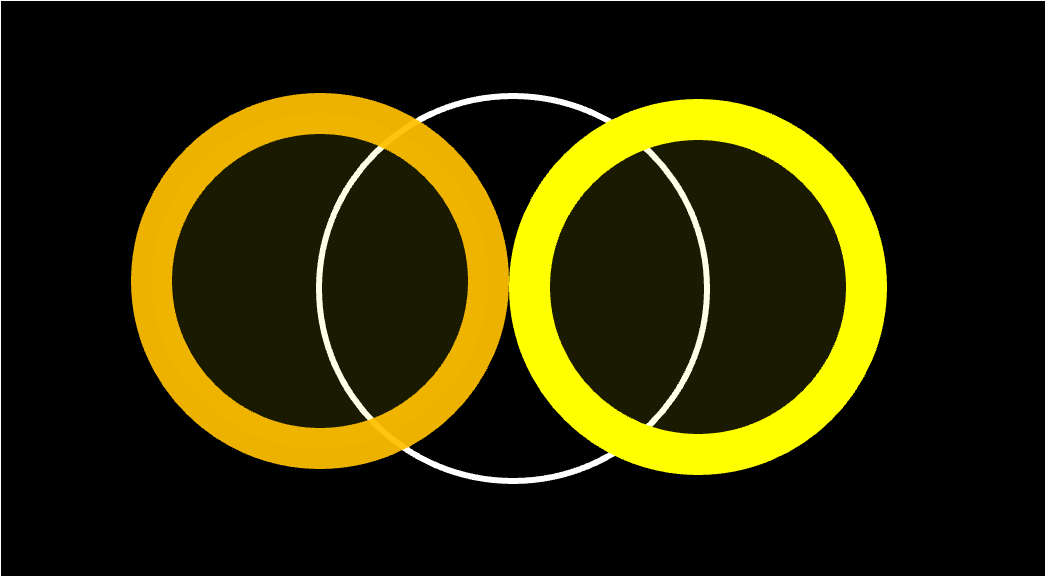} 
        \caption{In an ideal model, we show the expected components of skimming secondaries at 6 TeV:  prompt gamma-ray photons by neutral pions (a thin, white ring) and twin wider and deeper belts by  electron pairs decay secondaries of more penetrating muon pairs (yellow , negative, and orange positive, rings).}\label{fig:3}
    \end{minipage}\hfill
    \begin{minipage}{0.45\textwidth}
        \centering
        \includegraphics[width=0.8\textwidth]{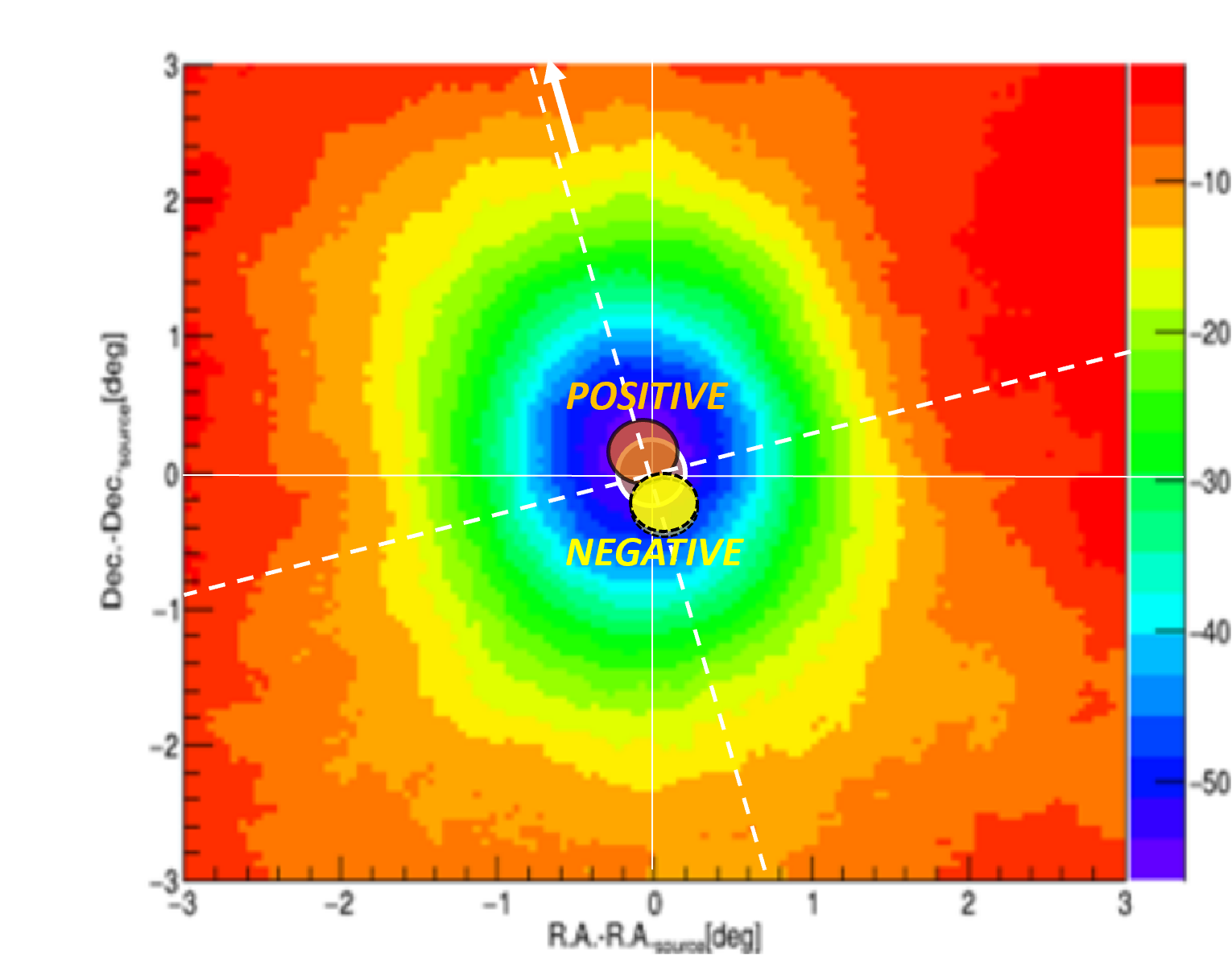} %
       \caption{The significance sky map of the Sun’s shadow as published by LHAASO\cite{Cui:2023e9} just before HAWC discovered the gamma-ray excess. %
       We overlap here the expected negative (yellow), neutral and positive (red) disk signals for the future data filtering.}
        \label{fig:5}
    \end{minipage}
\end{figure}

\begin{figure}[t]
\centering
\includegraphics[width=0.4\textwidth]{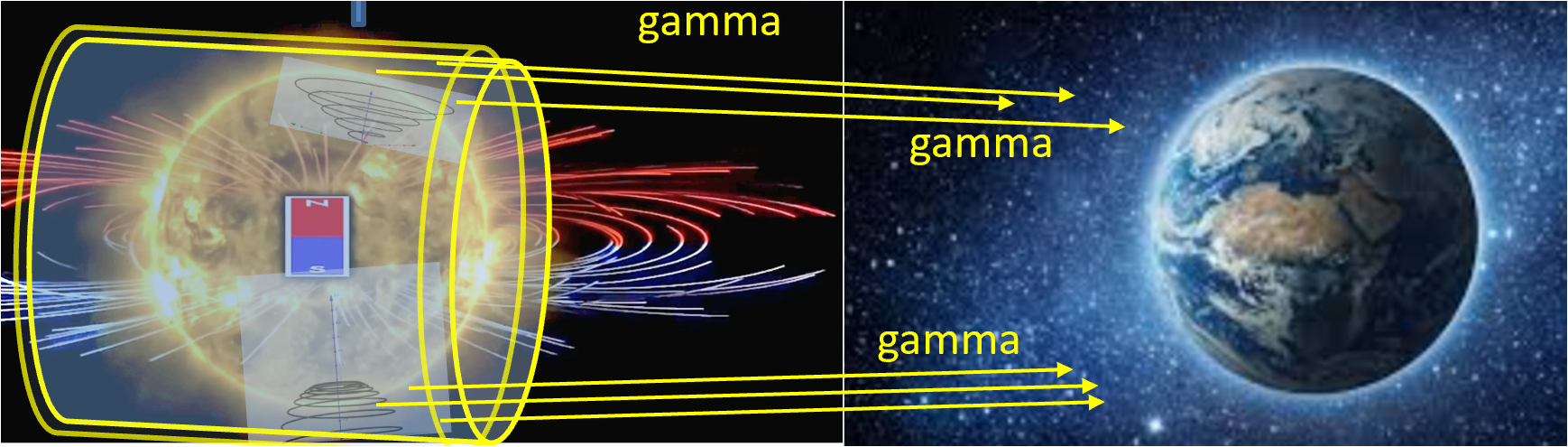} 
\caption{Schematic diagram showing CR skimming the Sun, forming a sort of tube: some TeV electron pairs are bent and captured in the upper solar atmosphere, spiraling along the solar radial magnetic lines. These TeV electron pairs could be feeding, via inverse Compton scattering, additional TeV gamma-ray photons directed toward Earth.}
\label{fig:8}
\end{figure}

\begin{figure}[b]
    \centering
    \begin{minipage}{0.45\textwidth}
        \centering
        \includegraphics[width=1.0\textwidth]{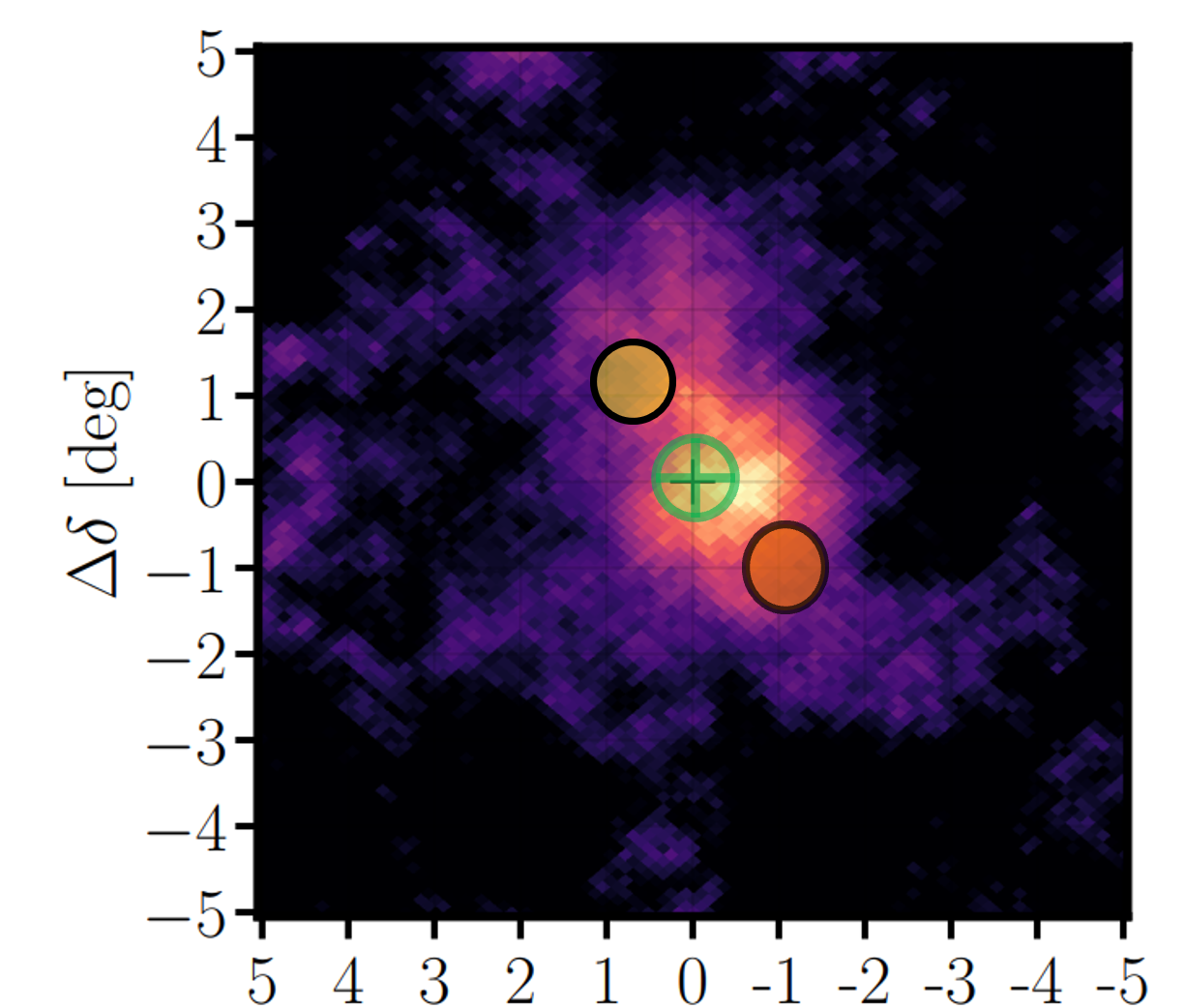} %
        \caption{The HAWC significance map of the gamma-ray signal within the Sun' shadow at the Solar minimum\cite{albert2023discovery} where we over-imposed the expected locations for the positive (red) and negative (yellow) shadows described in the text.}
        \label{fig:6}
    \end{minipage}\hfill
    \begin{minipage}{0.45\textwidth}
        \centering
        \includegraphics[width=0.9\textwidth]{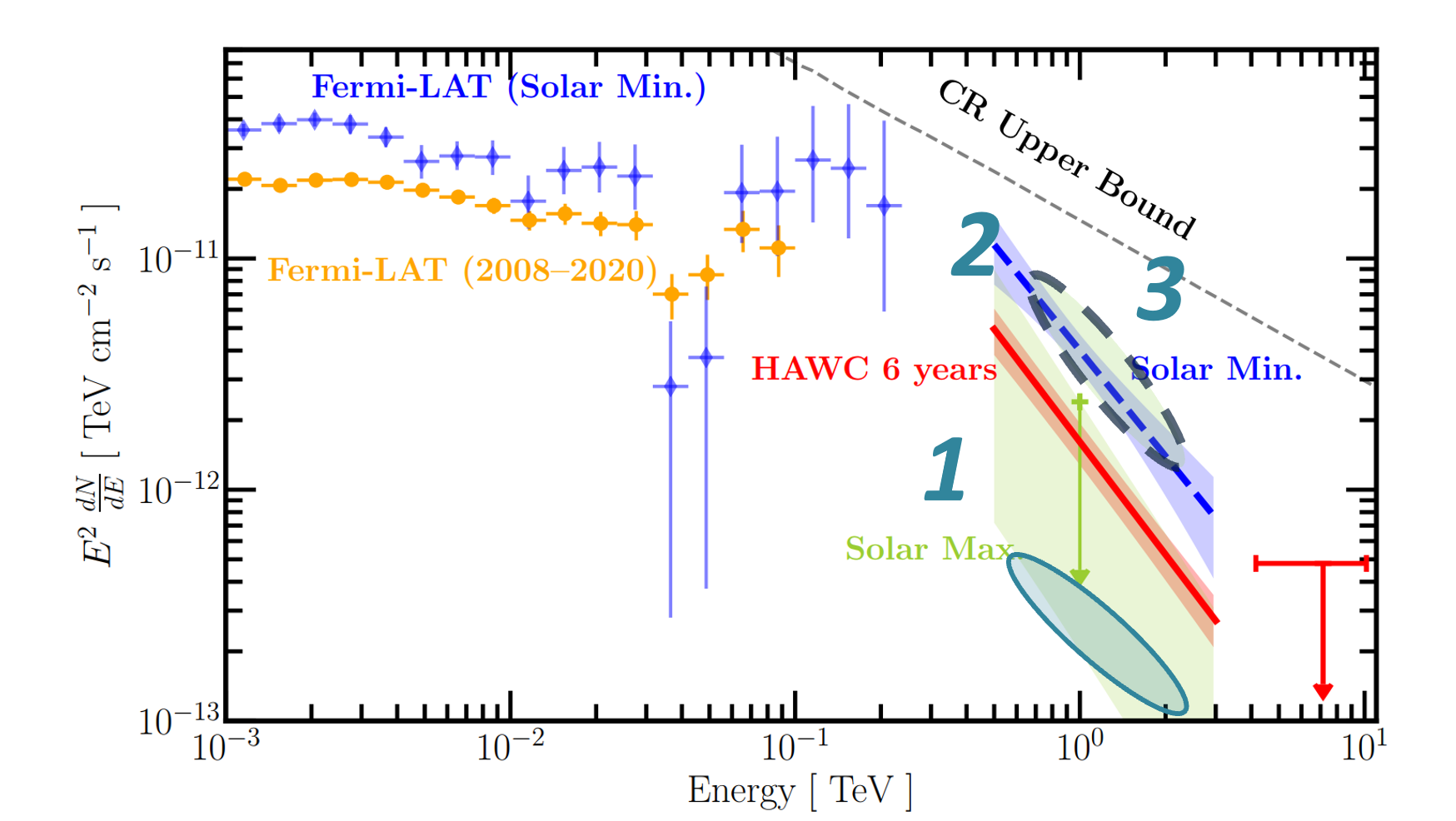} %
 \caption{The Energy Spectrum of the Sun, as seen by Fermi-LAT and by HAWC \cite{albert2023discovery}. We over-imposed  
 our earlier  expected muon-electron signals as calculated following  \cite{FargionICRC2023PoS1548} (lower green ellipse),  assuming only a thin gamma ring and a muon deeper penetration. A preliminary  new expected signal with an enhanced component of TeV muons assumed partially in coherent or random deflections by internal solar fields, (dashed curve).}
        \label{fig:7}
    \end{minipage}
\end{figure}

Let us also remind that  TeV   muons or electron pairs  (or any charged particle as proton), are bent by solar-planetary field by an approximated  angle $(1.3^o-1.5^o) (TeV/E_\mu $).   
Therefore, the twin gamma-like signals, the ideal positive-negative spots, will be smeared, but  located at opposite sides of the optical (neutral) solar disk. This ideal spots (not smeared) are described in figure \ref{fig:3}. The ones extracted from LHAASO data, before any signal filtering, are shown in figure \ref{fig:5} with the overlap ideal , gamma ray  and gamma-like component .   The ones extracted from HAWC data, after the signal filtering in order to disentangle gamma-ray and gamma-like , with our overlap of ideal gamma and gamma-like components, are shown in figure \ref{fig:6}.    The interesting count of these expected  charged spots signal, their numerical ratio respects the eventual neutral ones, is a  reading key of the inner process on solar atmosphere. The wider muon penetration and its beta decay in flight offer the most promising source of gamma-like signals. Other secondary emissions made both by local electron pairs bremsstrahlung, while skimming outside the solar atmosphere, or by ICS onto local optical  solar photons, while spiraling along radial solar lines,they are additional , but we believe minor,  components for the local gamma-ray signal from the Sun (see figure \ref{fig:8}).

\subsection{ Bending and turning  TeV electron or muons pairs on solar surface by local field}
TeV electron pairs may be bent  by  inner solar magnetic  fields, and  may shine also from different opening angles.  This bending   offers a wider emission areas of electron pairs (mostly the muon secondary ones).
The Larmor radius  $R_L$ at TeV energy in normal, Gauss like, field, is quite wide: 
 \begin{equation*} R_L = 33\cdot 10^3 ~\text{km} \cdot (E/TeV) \cdot (B/Gauss)^{-1}.\end{equation*} 
 The granularity of the solar atmosphere and the surface fields may bend  mostly the muons and partially electron pairs ( as we had shown, more opaque than the muons). Also electrons re-emission by bremsstrahlung or by  ICS on thermal Sun photons, may play a minor  role for gamma-rays local emission. The skimming electron pairs spiraling seconds or minute along solar magnetic lines (see Figure\ref{fig:4}) may  also re bounce solar photons as TeV photons.  The muon-electron bending, their up-down  splitting with respect to their solar tangent, may enlarge the muons skimming angle, offering  a  wider abundant gamma-like solar signal.    The ratio between the electron distance chord of   $D = 1.850 \cdot 10^3 \cdot  km $ with the TeV, the  Larmor radius shown above, is  at best (coherent bending)  nearly $3^o$ degree,  explaining a wider emission ring of prompt electron pairs from the Sun. Because the longer TeV muon survival distance $$ D = 6.2 \cdot 10^3 \cdot  km$$ and the same Larmor radius at same energy,  the muon  bending may be three times larger offering an open angle as large as $10^o$ degrees. A spread angle for  gamma-like sources from solar atmosphere, able to fit and tune with the over-abundant gamma-like events in HAWC. In our present  approximation, these two  smearing angles could  the keey to solve at best the HAWC    un-explained signals.

The ideal emission rings might be smeared by solar planetary spiral field. The 6 TeV energy in principle guarantees a partial overlap of the positive and negative spots, within the same optical one: for instance, see the LHAASO solar CR shadows at 6 TeV, with the expected twin bent spot of gamma-like muon-electron events (see   figure  \ref{fig:5}).

\section{Conclusions}

The high energy (TeV-PeV) CR and their secondaries hitting on the solar atmosphere had been revealed by the wide gamma-ray array HAWC around the Solar CR shadows  also with a  tiny smeared tail of {gamma-like} events.  Their discovery procedure is a very clever filtering of the hadrons from the electromagnetic events.
In short, they are able to extract from the entire 6 years full-sky dataset (almost 6 million events) a subset of 6300 gamma-ray events.
Such signals are spread at TeV within an angle of one to two degrees around the Sun.
  These more than a thousand per year over-abundant traces  observed by HAWC, should be soon revealed even at several or tens TeV  by the LHAASO, the most recent, widest km-size gamma-ray ground detector.  
Tens TeV muons may penetrate $1\%$ or $2\%$ of the solar height and a twice solar disk area. Unfortunately, CR spectrum at those energy is too diluted to fit the HAWC results. 
We have shown that, among the different CR  secondaries processes, skimming charged TeV muons, partially bent and smeared by the local solar magnetic field,  are decaying in their flight to us into electron pairs, offering the best candidate signals along a wide (5\%) of the solar ring area (see, for instance, figure \ref{fig:7} for past and new expected rates). Other processes as prompt $\pi^0$ decay in skimming gamma-rays, or other local gamma-ray photons ejection from the solar surface, are estimated to be quite  minor (about  $  1\%$ of the solar disk or below ten percent of the events).  These final charges exit and bending by interplanetary field  into a  twin electron pair flux should form also a  bipolar  gamma-like spot record,  around their  optical Solar shadows center.  Each component  (negative, positive, neutral) might be observed with better  resolution and statistic.   For instance, at  6 TeV the LHAASO  detection angular resolution is comparable to the solar solid angle, radius, $0.25^o$.  Thus, such a splitting  phenomena may be better disentangled and counted by a square kilometer array detector as LHAASO  (see figure  \ref{fig:5}).  The positive {gamma-like}  component will be mostly  over-lapping  with the same  hadron CR  solar shadows, while the negative one will be diluted in the  opposite side (see as HAWC map,  figure  \ref{fig:6}).  These signals would compete with the other \textit{"real"} local , un-deflected gamma-ray emission by the Sun.   Such thousands of events per year or  several tens of thousands of events per year,  respectively at HAWC  or,  hopefully,  soon at LHAASO,  could be  paving the way for the first gamma-ray and muon-electron pair astrophysics, as well as a new form of solar derived astronomy.    The ability to inspect along the inner solar atmosphere and solar fields using such muon radiography is relevant in itself.
Our main inspiring hope was \cite{fargion2018signals} (and still is\cite{FargionICRC2023PoS1548}) that such a new  powerful tool based on muons decay in flight  might be also extended to the lunar shadows.  Indeed, the Moon has no atmosphere, nor much CR skimming secondaries as Sun or Earth at horizons.  No  atmospheric CR noise muons comes from the Moon.  But, surprisingly, ten or tens of TeV  (up  to  nearly $62$  TeV ) astronomical muon neutrino might be  interact on  lunar soil facing the Earth.  Their out-going muons may emerge from the Moon to us,  from a region of the sky unaffected by the huge CR Earth's atmospheric noise, decaying in flight as electron pairs.   Rarely, with a probability of about $10^{-4}$ -- $10^{-3}$, incoming   {astrophysical } muon neutrinos , can do it: their secondary muons of  few or tens of TeV can escape the external  lunar crust (a few or ten  kilometers of the lunar rock crust depth) and head towards us as muon and later  as electron pairs, reproducing  the legacy similar to that of solar {gamma-like}  observable terrestrial air-showers on HAWC or LHAASO.
They could be  leading to a much  rare gamma-like twin, positive and negative, discrete events around the same Moon CR shadows.   The dream for  such a new muon  gamma-like neutrino astronomy (based on a final electron air-showers coming from muon ejected from the inside of Moon shadows), recalls the   similar tau air-shower neutrino phenomenon on Earth, a possible discover tool often referred as skimming tau neutrino astronomy.  Likewise, Muon neutrinos from the Moon could play a comparable role in the pair muon neutrino astronomy, thanks to their rare  TeV electrons  pairs  and the present widest LHAASO array.

\section*{Acknowledgements}
 \noindent The research of D.S. was carried out in the Southern Federal University with financial support of the Ministry of Science and Higher Education of the Russian Federation {\small(State contract GZ0110/23-10-IF)}. 

\bibliographystyle{JHEP}
\bibliography{daf2025v25}

\end{document}